\documentstyle[12pt,epsfig]{article}
\begin{document}
\addtocounter{footnote}{1}
\title{
Unification of Dark Matter and Dark Energy:
 the Inhomogeneous Chaplygin Gas
 }
\author{Neven Bili\'{c}\thanks{Permanent
 address:
Rudjer Bo\v skovi\'c Institute,
P.O. Box 180, 10002 Zagreb, Croatia;
 \hspace*{5mm} Email: bilic@thphys.irb.hr}\ ,
Gary B.\ Tupper, and Raoul D.\ Viollier\thanks{
 Email: viollier@physci.uct.ac.za}
\\
Institute of Theoretical Physics and Astrophysics, \\
 Department of Physics, University of Cape Town,  \\
 Private Bag, Rondebosch 7701, South Africa \\
 }
\maketitle
\begin{abstract}
We extend the world model of Kamenshchik et al.\
to large perturbations by formulating a
Zeldovich-like approximation.
We sketch how this model unifies dark matter with
dark energy in a geometric setting
reminiscent of $M$-theory.
\end{abstract}
%
%

After nearly two decades of reign \cite{kol1}, the Einstein-de Sitter
dust model has been swept aside by observations of high redshift
 supernovae \cite{perl2} which suggest
that the Hubble expansion is
accelerating.
When combined with the Boomerang/Maxima data
\cite{deb3}
showing
that the location of the first acoustic peak in
the power spectrum of
the microwave
background is consistent
with the inflationary prediction $\Omega=1$, the
evidence for a net equation of state of the cosmic
fluid lying in the range $- 1 \leq w = P/\rho < - 1/3$
is compelling.
 Parametrically,
$w=P_{\rm DE}/(\rho_{\rm DE}+\rho_{\rm DM})=-\Omega_{\Lambda}$
 gives a ratio of unclustered
dark energy to clustered dark matter of order 7:3, thereby also resolving
the longstanding $\Omega_{\rm DM} < 1$ puzzle \cite{kol1,peeb4} implied
by peculiar velocity fields.
The theoretical implications
of these dicoveries
 are profound.
Simply appending a nonzero
cosmological constant $\Lambda$ to standard cold dark matter
($\Lambda$CDM, \cite{peeb4}) invites
anthropic arguments \cite{car5} as to why
both
$\Omega_{\rm DM}$ and
$\Omega_{\Lambda} $ are of order unity today.
False vacuum models
have been proposed \cite{dva6},
and, in one formulation \cite{bar7}, linked to
quantum gravity effects and axionic CDM.

Currently, the most
popular alternative is Quint\-es\-sen\-ce \cite{peeb8} which conventionally
entails a real, homogeneous scalar field
of mass $m_{Q} < H_{0}$ whose potential
is arranged such that it silently tracks radiation and matter,
gracefully
entering a dominant de Sitter phase with an intermediate mixed
epoch (QCDM) today.
Acceptable
models can be obtained using pseudo-Goldstone
bosons \cite{cob9}, while
spintessence \cite{boy10} is a twist on this theme
involving a complex quintessence field.

If $\Lambda$CDM appears
too coincidental,
QCDM needs two distinct fields,
 one to describe dark matter, the other dark energy.
Economy would suggest that
 dark energy and dark matter should be different
manifestations of the same entity.
Wetterich \cite{wet11} has speculated that dark
matter may be cosmon (quintessence) lumps while Kasuya \cite{kas12}
has pointed out that spintessence-like models are generally unstable to
formation of Q-balls which behave as pressureless matter. Because they
rely upon the coupled nonlinear
scalar-gravitational fields even into the
nonlinear regime,
it is difficult to assess these proposals in the absence
of large-scale numerical simulations.
This is in marked contrast to
$\Omega_{\rm CDM}$ = 1 where the Zeldovich approximation \cite{zel13}
allows one to obtain an intuitive picture.
Being essentially a variant of
$\Lambda$CDM, the Barr-Sechel scenario \cite{bar7} does not suffer
this criticism.

In a recent paper Kamenshchik,
Moschella and Pas\-qui\-er \cite{kam1}
have studied a homogeneous model based on a single fluid obeying the
Chaplygin gas equation of state
\begin{equation}
P = -A/\rho  ,
\label{eq001}
\end{equation}
which has been intensively
investigated for its solubility in 1+1-dimensional space-time,
its supersymmetric
extension and connection to d-branes \cite{jac1}.
Implementing this in
the relativistic energy conservation equation,
the density evolves according to
\begin{equation}
\rho (a) = \sqrt{A + B/a^{6}}\, ,
\label{eq002}
\end{equation}
where $a$ is the scale factor and $B$ an integration constant.
This model smoothly interpolates between dust and de Sitter phases
without ad hoc assumption.
Fabris, Goncalves and de Souza \cite{fab16}
have endeavoured to examine density perturbations to this model,
but, owing to an unfortunate choice of the lightcone gauge,
their unperturbed
Newtonian equations cannot reproduce (\ref{eq002}), which is a
prerequisite.
In any case, the universe
is very inhomogeneous \cite{wu17} today, and if one takes this idea
seriously
one needs a Zeldovich-like nonperturbative approach
adopted to the case $P \neq 0$.

The purpose of
this letter is to give such a formulation covariantly
 and in sufficient
generality that it can be adopted to any balometric or parametric
equation of state.
Consider the
Lagrangian
\begin{equation}
{\cal{L}}
 =  g^{\mu \nu} {\Phi^*}_{, \mu}
\Phi_{, \nu}
 -V(|\Phi|^{2})
\label{eq101}
\end{equation}
for  a complex scalar field $\Phi = (\phi/\sqrt{2}\, m)
\exp (-i m \theta)$, where $m$ is the mass appearing in the
potential $V$.
The Lagrangian (\ref{eq101}), expressed in terms of
$\phi$ and $\theta$, reads
\begin{equation}
{\cal{L}}
 = \frac{1}{2} g^{\mu \nu}\left(\phi^2
 \theta_{,\mu} \theta_{,\nu}
+\frac{1}{m^2}
 \phi_{,\mu}\phi_{,\nu}\right)
 -V(\phi^2/2) .
\label{eq102}
\end{equation}
We now apply  the
 Thomas-Fermi
approximation \cite{par18}.
In contrast to Kamenshchik et al.\ \cite{kam1}
who assumed spatial
homogeneity,
we allow for
space-time variations of
$\phi$ on scales
   larger than  $m^{-1}$.
More precisely, we assume that
\begin{equation}
\phi_{, \mu} \ll m \phi ,
\label{eq105}
\end{equation}
and  hence, we may neglect the second term
in the brackets on the right hand side of
 Eq.\ (\ref{eq102}).
This yields  the  Lagrangian
\begin{equation}
{\cal{L}}_{\rm
TF} = \frac{\phi^{2}}{2} g^{\mu \nu} \theta_{, \mu}
\theta_{, \nu} - V (\phi^{2}/2)
\label{eq003}
\end{equation}
 and the equations of motion for the fields
$\phi$ and $\theta$
\begin{equation}
g^{\mu \nu} \theta_{,\mu} \theta_{,\nu} = V'(\phi^2/2)\, ,
\label{eq004}
\end{equation}
\begin{equation}
(\sqrt{-g}\,\phi^2
 g^{\mu \nu} \theta_{,\nu} )_{,\mu}=0 ,
\label{eq103}
\end{equation}
where $V'(x)=dV/dx$.
Assuming $V' >0$, the field $\theta$ may be treated as a
velocity potential for the fluid 4-velocity
\begin{equation}
U^{\mu} = g^{\mu \nu} \theta_{,\nu} / \sqrt{V'} \, ,
\label{eq005}
\end{equation}
that is restricted to the mass shell, i.e.,
$U_{\mu} U^{\mu}$ = 1.
As a consequence,
 the stress-energy tensor $T^{\mu \nu}$ constructed
from the Lagrangian (\ref{eq003})
takes the perfect fluid
form, with the parametric equation
of state
\begin{equation}
\rho = \frac{\phi^{2}}{2} V'+ V ,
 \hspace{1cm} P = \frac{\phi^{2}}{2}
V' - V  .
\label{eq006}
\end{equation}
This procedure may be reversed
as in the case of Newtonian limit \cite{bil}.
Suppose
the equation of state is given, e.g.,
in parametric form $P (\psi), \rho (\psi)$.
From Eqs.\ (\ref{eq006}) we find
\begin{equation}
\ln (\phi^{2}) = \int \frac{d \rho -dP}{\rho+P} \, .
\label{eq007}
\end{equation}
Using this
one obtains $\rho (\phi^{2})$ or
$\psi (\phi^{2})$ and hence the potential
\begin{equation}
V = \frac{1}{2} (\rho - P)
\label{eq008}
\end{equation}
of the field theory associated with the fluid,
the integration constant being
fixed by the  $P=0$ limit.
For the Chaplygin gas one  finds
$\rho=\phi^2$ and
\begin{equation}
V = \frac{1}{2} \left(\phi^{2} + \frac{A}{\phi^{2}}\right).
\label{eq009}
\end{equation}
We remark that this potential is very different from
that of Kamenshchik et al.\ \cite{kam1} whose
derivation is based on the
assumption of homogeneity.
Note that (\ref{eq007}) has the structure of
a renormalization group for the fluid-field map.

For a sensible theory with  $ 0 \geq w \geq - 1$
and a speed of sound satisfying
 $0\leq C_{s}^{2} = d P/d
\rho \leq 1 $, the relativistic
limits of these inequalities should coincide,
which uniquely selects
(\ref{eq001}).
It is interesting to note that $P = - A/ \rho^{\alpha}$ gives
the functional flow
\begin{equation}
\phi^2(\rho)=\rho^{\alpha}
(\rho^{\alpha +1}-A)^{\frac{1-\alpha}{1+\alpha}}
\label{eq301}
\end{equation}
which, as $\rho^{\alpha +1}-A
\rightarrow 0^{+}$, resembles that of the periodic gaussian model
\cite{kog19}.
The critical line $\phi^{2} = \rho$
describes
again the Chaplygin gas
and $V^*=(\phi^2+A/\phi^2)/2$
is the corresponding critical potential.
Introducing $K = \phi^{2} / \sqrt{A}$ which
measures the dimensionless coupling of the $\theta$
field in (\ref{eq003}), the potential $V^{*}$ is self-dual
under $K \rightarrow K^{-1}$, reminiscent of M-theory string dualities
\cite{pol20}.
Indeed, the physical content of the full ${\cal{L}^{*}}$ shares
this duality,
as can be seen through eliminating $\phi^{2}$ by its algebraic
equation of motion.
This yields a Born-Infield theory
\cite{jac2}
\begin{equation}
{\cal{L}^{*}} = - \sqrt{A}\,
\sqrt{1-g^{\mu\nu}\theta_{,\mu}\theta_{,\nu}}
\label{eq010}
\end{equation}
describing
space-time as a 3+1 brane in a 4+1 dimensional bulk \cite{sun}.
The velocity potential $\theta$ in (\ref{eq005})
measures surface excursion in
the fifth dimension in Eq.\ (\ref{eq010}).

Notably, the Chaplygin gas appears in the stabilization of branes in
Schwarzschild-AdS black
hole bulks as a critical theory at the horizon \cite{kam2}.
Also, a free-energy similar to Eq.\ (\ref{eq009}) occurs in
the stringy analysis of black holes in three dimensions \cite{ram24}.
Replacing Rama's ``string bits" with ``brane cells", the free energy
(\ref{eq009}) follows on dimensional grounds,
$\phi^{-1}$ corresponding to the
size of the system.

Of course the primary question is whether
${\cal{L}^{*}}$ yields a reasonable
inhomogeneous cosmology.
For this purpose we  do not eliminate $\phi$
from the Lagrangian (\ref{eq003}) with (\ref{eq009}), writing
the field equation (\ref{eq103}) in the form
\begin{equation}
\left( \sqrt{-g \rho (\rho + P)}\, U^{\mu} \right)_{, \mu} = 0
\label{eq011}
\end{equation}
instead.
To solve this equation, it is convenient to use
comoving coordinates.
In the general comoving coordinate system the
4-velocity vector is given by
$U^{\mu}=\delta^{\mu}_0/\sqrt{g_{00}}\,$.
 Eq.\ (\ref{eq011}) then becomes
\begin{equation}
\left(\sqrt{\frac{-g}{g_{00}}(\rho^2-A)}\:\right)_{,0} = 0,
\label{eq106}
\end{equation}
with the solution
\begin{equation}
\frac{-g}{g_{00}}(\rho^2-A) = B(\mbox{\boldmath $x$}),
\label{eq107}
\end{equation}
where $B$ is an arbitrary function independent of $x^0$.
For a general metric $g_{\mu\nu}$, the proper time is
$d\tau= \sqrt{g_{00}}\, dx^0$ and
$-g/g_{00}\equiv \gamma$ is the determinant of the
induced 3-metric
\begin{equation}
\gamma_{ij} =\frac{g_{i0}g_{j0}}{g_{00}}-g_{ij} \, .
\label{eq108}
\end{equation}
Noting that for the relevant scales $B(\mbox{\boldmath $x$})$
can be considered to be
 approximately
constant, we obtain
\begin{equation}
\rho = \sqrt{A + \frac{B}{\gamma}}
\label{eq012}
\end{equation}
as the generalization of
Eq.\ (\ref{eq002}),
in synchronous coordinates $(t=\tau,\mbox{\boldmath $x$})$.
We stress here a peculiarity of
negative versus ordinary positive pressure
fluids: for the former,
as the fluid becomes ultrarelativistic,
 the weak
energy condition $\rho+P\geq 0$
is saturated
so that ${T^{\mu\nu}}_{;\nu}=0$ implies
$P_{,\mu}=0$ and  there are no pressure gradients.
Thus in both the nonrelativistic  and ultrarelativistic
regimes the synchronous coordinates are comoving.

Based on this observation,
we can take
over from the dust case \cite{mat25}
the geometric implementation of the Zeldovich approximation as a map
from Lagrange ({\boldmath $q$}) to Euler (comoving-synchronous,
{\boldmath $x$}) coordinates   inducing
the 3-metric
\begin{equation}
\gamma_{ij}=\delta_{kl} {D_i}^k
{D_j}^l .
\label{eq013}
\end{equation}
Here $D$ is the deformation tensor
\begin{equation}
{D_i}^j=a(t)\left({\delta_i}^j-b(t)
\frac{\partial^{2}\varphi (\mbox{\boldmath $q$})}{\partial q^{i}
\partial q_{j}} \right),
\label{eq014}
\end{equation}
and $\varphi$ the peculiar velocity potential.
The function $b$,
which describes the evolution of the density contrast,
may be calculated in the standard way
\cite{pee} treating
the quantity
\begin{equation}
h=2b(t)
{\varphi_{,i}}^i
\label{eq201}
\end{equation}
as a perturbation.
From Eqs.\ (\ref{eq012})-(\ref{eq201}) one can derive
\begin{equation}
\rho\simeq\bar{\rho}(1+\delta), \;\;\;\;
P\simeq-\frac{A}{\bar{\rho}}(1-\delta),
\label{eq202}
\end{equation}
with
\begin{equation}
\bar{\rho}=\sqrt{A+B/a^6}\,, \;\;\;\;
\delta=\frac{h}{2}(1+w),
\label{eq203}
\end{equation}
where
\begin{equation}
w=-\frac{A}{\bar{\rho}^2}.
\label{eq204}
\end{equation}
Using these equations
and the synchronous metric
defined in Eq.\ (\ref{eq013})
with (\ref{eq014})
we obtain
the 0-0 component of the Einstein field equations in the form
\begin{equation}
-3\frac{\ddot{a}}{a}+
\frac{1}{2}\ddot{h}+
\frac{\dot{a}}{a}\dot{h}=
4\pi \bar{\rho} [(1+3w)+(1-3w)\delta].
\label{eq205}
\end{equation}
The unperturbed part of this equation
\begin{equation}
\frac{\ddot{a}}{a}=
-\frac{4\pi}{3} \bar{\rho} (1+3w)
\label{eq206}
\end{equation}
coincides with the  second Friedmann equation.
The first-order part of (\ref{eq205}) may be written as a
diferential equation for $b(a)$, i.e.,
\begin{equation}
\frac{2}{3} a^{2} b'' + a (1 - w) b' =
(1 + w) (1 - 3 w) b ,
\label{eq015}
\end{equation}
where the prime denotes the derivative with respect to $a$.
The function  $w(a)$ may be conveniently expressed in terms
of the standard parameter
$\Omega_{\Lambda}$
which was
introduced previously, i.e.,
\begin{equation}
w(a) = - \frac{\Omega_{\Lambda} a^{6}}{1-\Omega_{\Lambda}+
\Omega_{\Lambda} a^{6}}.
\label{eq016}
\end{equation}
\begin{figure}[t]
\centering
\epsfig{file=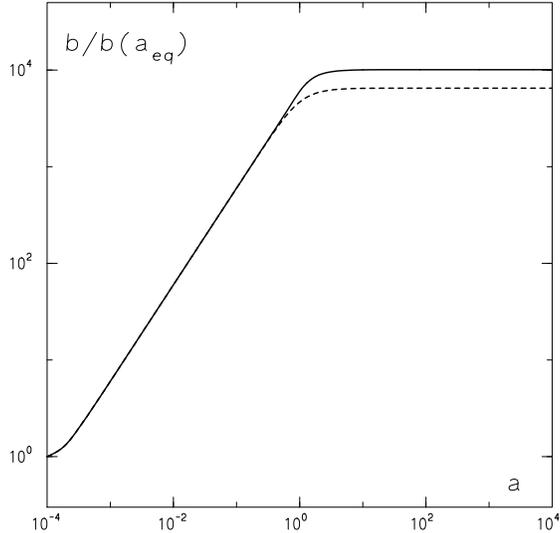,height=10cm}
\caption{
 Evolution of $b (a)/ b(a_{\rm eq})$
from $a_{\rm eq}= 1.0 \times 10^{-4}$ for $\Omega_{\Lambda} = 0.7$
    and $b' (a_{\rm eq}) = 0$,
   for the Chaplygin gas (solid line) and $\Lambda$CDM (dashed line).
}
\label{fig1}
\end{figure}

In Fig.\ \ref{fig1}
the evolution of $b(a)$ is shown from radiation-matter
equality with
the initial condition $b'(a_{\rm eq})=0$.
Owing to both the high power of $a$ appearing in
Eq.\ (\ref{eq016}) and
the factor of (1 - 3$w$) in (\ref{eq015}),
the difference
to the standard CDM scenario at $a$ = 1, i.e., today,
is negligable.
Note that, while the metric perturbation
$h\propto b$
saturates,
the density contrast
$\delta \propto (1+w)b$ vanishes
for large $a$.
In the same figure we give the corresponding result
for $\Lambda$CDM which is obtained by omitting
the factor (1-3$w$) in Eq.\
(\ref{eq015}) and changing $a^{6}$ to $a^{3}$ in Eq.\
(\ref{eq016});
the contrast $\delta$
in that case is 50\% smaller at $a$ = 1.
Thus for the inhomogeneous
Chaplygin gas we can take over the dust results
bodily up to $z \simeq$ 0.
The picture which emerges is that on caustics,
where galactic halos and clusters form,
we have
$w \simeq$ 0, i.e., the fluid behaves as dark
matter.
Conversely, in the voids $w\simeq-\Omega_{\Lambda}$
 drives the
acceleration as dark energy.
Here the answer to the coincidence question
mentioned earlier is
that acceleration sets in only once
the observed cellular structure develops.

In conclusion, we have shown that the inhomogeneous Chaplygin gas offers a
simple unified model of dark matter and dark energy.
It may be worth pointing out
that the potential $V$ obtained from the simple equation of state
(\ref{eq001}) may easily be generalized by adding a more
complicated interaction term, e.g., a power of $\phi$
greater than 2, and in this way generate a more refined equation
of state for dark matter, instead of $w=0$.
This would not alter our general picture, as long as
this additional interaction term is  much smaller
than $\sqrt{A}$ when $\phi^2$ approaches $\sqrt{A}$.
In particular, one can add a thermal component to
(\ref{eq001}) which,
although irrelevant to the large-scale structure
and acceleration, provides a finite phase-space density
inferred from galactic cores
\cite{bil2}.
Ordinary matter can be included using
Sundrum's \cite{sun} effective field
theory; the $\theta$ field is derivatively coupled and thus harmless.
The fact that this unification is achieved
in a geometric framework,
having roots in the  ``theory of everything", makes
this scenario all the more remarkable.

We thank
J.C.\ Fabris, S.B.V.\ Goncalves, P.E.\ de Souza,
and C.\ Wetterich
for valuable  comments.
This
research is in part supported by the Foundation of Fundamental
Research (FFR) grant number PHY99-01241 and the Research Committee of
the University of Cape Town.
The work of N.B.\ is supported in part by
the Ministry of Science and Technology of the Republic of Croatia
under Contract No.\ 00980102.

\end{document}